# On the $k_\perp$ dependent gluon density of the proton

Johannes Blümlein

*DESY–Zeuthen, Platanenallee 6, D–15735 Zeuthen, Germany*

**Abstract**

The $k_\perp$ dependent gluon distribution is calculated accounting for the resummation of small $x$ effects due to the Lipatov equation. It is represented by a convolution of a gluon density in the collinear limit and a universal function $\mathcal{G}(x, k^2, \mu)$ for which an analytic expression is derived.

# On the $k_\perp$ dependent gluon density of the proton

Johannes Blümlein

DESY–Zeuthen, Platanenallee 6, D–15735 Zeuthen, Germany

**Abstract**

The $k_\perp$ dependent gluon distribution is calculated numerically accounting for the resummation of small $x$ effects due to the Lipatov equation. It is represented by a convolution of a gluon density in the collinear limit and a universal function $\mathcal{G}(x, k^2, \mu)$ for which an analytic expression is derived.

## 1. Introduction

In the small $x$ range new dynamical effects are expected to determine the behaviour of structure functions. The evolution of parton densities is effected by terms due to non strong $k_\perp$ ordering and eventually by screening contributions. A description of contributions of this type requires to generalize the factorization of the hadronic matrix elements into coefficient functions and parton densities in which the $k_\perp$ depencence is not integrated out [1].

This factorization covers the case of collinear factorization in the limit that the $k^2$ dependence of the coefficient function is neglected. The $k_\perp$ dependent gluon density accounts for the resummation of small $x$ effects. In the present paper we will consider those due to the Lipatov equation only. Since this equation behaves infrared finite no other singularities will emerge than in the case of mass factorization. The collinear singularities are delt with in the same way in the case of $k_\perp$ factorization.

In the present paper the $k_\perp$ dependent gluon distribution is calculated for the case of the scheme [2, 3]. It can be represented as the convolution of the gluon density in the collinear limit $g(x, \mu)$ and a function $\mathcal{G}(x, k^2, \mu)$ for which an analytic expression will be derived. A numerical comparison of the $k_\perp$ dependent gluon densities accounting for the solution of the Lipatov equation and in the double logarithmic approximation (DLA) is given.

## 2. $k_\perp$ Factorization and the $k_\perp$ dependent gluon distribution

The factorization relation for an observable $O_i(x, \mu)$ reads

$$O_i(x,\mu) = \int dk^2 \hat{\sigma}_{O_i}(x, k^2, \mu) \otimes \Phi(x, k^2, \mu) \quad (1)$$

where $\hat{\sigma}_{O_i}(x, k^2, \mu)$ and $\Phi(x, k^2, \mu)$ denote the $k^2$ dependent coefficient function and parton density♯, respectively. Eq. (1) can be rewritten as [2, 3]

$$O_i(x,\mu) = \hat{\sigma}^0_{O_i}(x,\mu) \otimes G(x,\mu) \\ + \int_0^\infty dk^2 \left[\hat{\sigma}_{O_i}(x,k^2,\mu) - \hat{\sigma}^0_{O_i}(x,\mu)\right] \Phi(x,k^2,\mu) \quad (2)$$

with $\hat{\sigma}^0_{O_i}(x,\mu) = \lim_{k^2 \to 0} \hat{\sigma}_{O_i}(x, k^2, \mu)$. The first addend in (2) describes the conventional contribution due to collinear factorization. The second term contains the new contributions. Note that $\Phi(x, k^2, \mu)$ starts with terms $\propto \overline{\alpha}_s$. It has therefore *not* the interpretation of a probability density and may even become negative.

As shown in [2] the $k_\perp$ dependent gluon distribution associated to eq. (2) reads in moment space

$$\widetilde{\Phi}(j, k^2, \mu) = \gamma_c(j, \overline{\alpha}_s) \frac{1}{k^2} \left(\frac{k^2}{\mu^2}\right)^{\gamma_c(j, \overline{\alpha}_s)} \widetilde{g}(j, \mu) \quad (3)$$

where $\mu$ denotes a factorization scale, $\overline{\alpha}_s = N_c \alpha_s(\mu)/\pi$, and $g(x, \mu)$ is the gluon density. Eq. (3) accounts for the small $x$ behaviour due to the Lipatov equation. Here

♯ We will consider the gluon density in the present paper only.

$$j-1 = \overline{\alpha}_s \chi(\gamma_c(j, \overline{\alpha}_s)), \quad \chi(\gamma) = 2\psi(1) - \psi(\gamma) - \psi(1-\gamma). \tag{4}$$

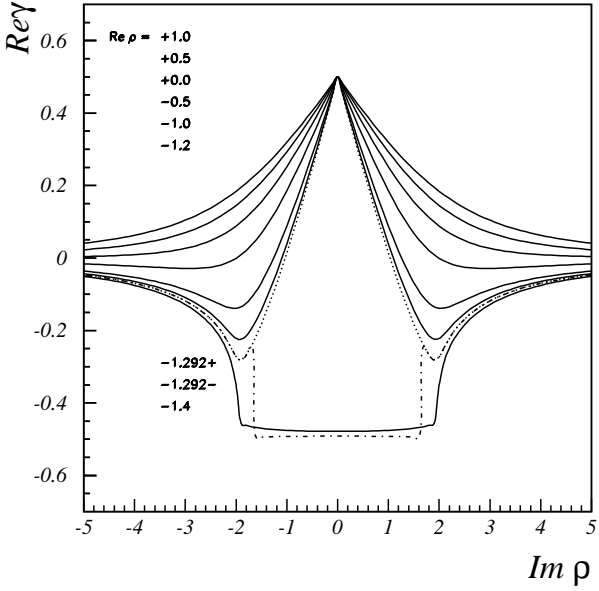

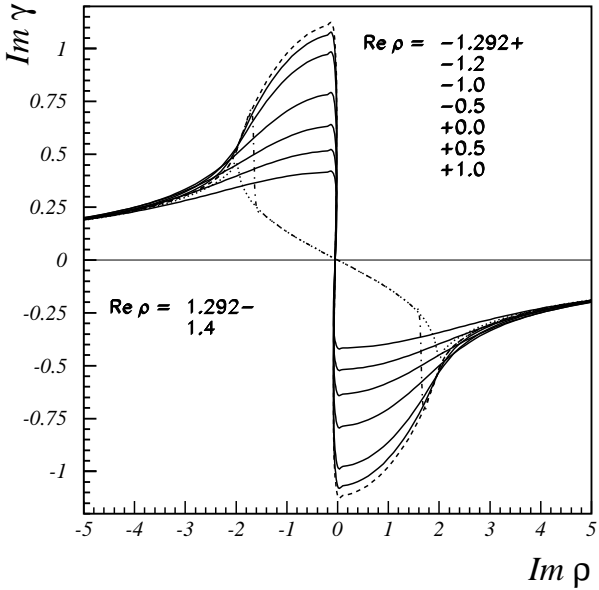

**Fig. 1** Real and imaginary part of $\gamma_c$ vs $Re(\rho)$ and $Im(\rho)$.

In $x$ space the $k_\perp$ dependent distribution is given by

††In several recent approaches [4] to describe $k_\perp$ dependent gluon distributions phenomenological Ansätze were used based on solutions of *inhomogeneous* Lipatov equations. Note that these descriptions are not related to eq. (2) and [2, 3].

$$\Phi(x, k^2, \mu) = \mathcal{G}(x, k^2, \mu) \otimes g(x, \mu), \tag{5}$$

correspondingly, with

$$\int_0^{\mu^2} dk^2 \Phi(x, k^2, \mu) = \delta(1-x). \tag{6}$$

The function $\mathcal{G}(x, k^2, \mu)$ is universal and can be calculated numerically by a contour integral in the complex plane over the first factor in eq. (3). Since the solution of eq. (4) is multivalued the Mellin inversion to $x$ space requires to select the branch in which for asymptotic values of $j \in \mathcal{C}$ $\gamma_c$ approaches the perturbative result $\gamma_c(j, \overline{\alpha}_s) \sim \overline{\alpha}_s/(j-1)$ for small values of $\overline{\alpha}_s$.

We solved eq. (4) under this condition numerically using an adaptive Newton algorithm. The solution is characterized by the well-known branch point at $\rho \equiv (j-1)/\overline{\alpha}_s = 4\ln 2$ and two further conjugate branch points (cf. also [5]). As shown in figure 1 the 'ridge' in the real part due to the singularity at $\rho = 4\ln 2$ persists until $Re\rho \sim -1.3$ and turns into a flat form with $Re(\gamma_c) \sim -0.5$ for $|Im\rho| < 1.5$. At the same time the imaginary part of $\gamma_c$ becomes continuous again.

## 3. An analytical solution for $\mathcal{G}(x, k^2, \mu)$

Because the integration contour has to be situated outside the range of the singularities of $\gamma_c$ one may expand $\gamma_c(j, \overline{\alpha}_s)$ into a Laurent series over $\rho$

$$\gamma_c(j, \overline{\alpha}_s) = \sum_{l=1}^\infty g_l \rho^{-l} \tag{7}$$

The coefficients $g_l$ are given in [6] up to $l = 20$ in analytical form extending an earlier result [7]. For small values of $|Im\rho|$ ($|Im\rho| < 2$) the truncated Laurent series leads to an oscillatory behaviour and eq. (7) does no longer serve to be an appropriate description of $\gamma_c$ (cf. [6]).

Using (7) a corresponding expansion may be performed for

$$k^2 \widetilde{\mathcal{G}}(j, k^2, \mu) = \gamma_c(j, \overline{\alpha}_s) \exp[\gamma_c(j, \overline{\alpha}_s) L] \tag{8}$$

with $L = \ln(k^2/\mu^2)$. For the single terms of the Laurent series in $\rho$ the Mellin transform can be carried out analytically. Here it is important to expand the exponential in eq. (8) in such a way that the lowest order term in $\overline{\alpha}_s$ of $\gamma_c$ is kept in exponential form. One obtains

$$k^2 \mathcal{G}(x, k^2, \mu) = \frac{\overline{\alpha}_s}{x} I_0\left(2\sqrt{\overline{\alpha}_s \log(1/x) L}\right)$$

$$\times \quad I_{\nu-1}\left(2\sqrt{\overline{\alpha}_s \log(1/x)L}\right), L > 0. \quad (9)$$

The coefficients $d_\nu(L)$ are given in ref. [6]. Up to $\nu = 20$ they contain at most terms $\propto L^4$. The first term in eq. (9) denotes the Green's function in DLA.

For $L \to 0$ (9) takes the form

$$k^2 \mathcal{G}(x, k^2, \mu) = \frac{\overline{\alpha}_s}{x} \sum_{l=1}^{\infty} \frac{g_l}{(l-1)!} \left[\overline{\alpha}_s \left(\frac{1}{x}\right)\right]^{l-1}, \quad (10)$$

and for $L < 0$ (i.e. $k^2 < \mu^2$) one has

$$k^2 \mathcal{G}(x, k^2, \mu) = \frac{\overline{\alpha}_s}{x} J_0\left(2\sqrt{\overline{\alpha}_s \log(1/x)|L|}\right)$$
$$+ \frac{\overline{\alpha}_s}{x} \sum_{\nu=4}^{\infty} d_\nu(L) \left(\frac{\overline{\alpha}_s \log(1/x)}{|L|}\right)^{(\nu-1)/2}$$
$$\times \quad J_{\nu-1}\left(2\sqrt{\overline{\alpha}_s \log(1/x)|L|}\right). \quad (11)$$

Thus for $k^2 \to 0$ damped, oscillating modes are obtained which vanish faster than $1/|L|^{-1/4}$.

## 4. Numerical Results

The $k_\perp$ dependent gluon distribution $\Phi(x, K^2, \mu)$ (scaled by $k^2$) is shown in figure 2 as a function of $x$ and $k^2$ for $\mu^2 = 20\,\text{GeV}^2$.

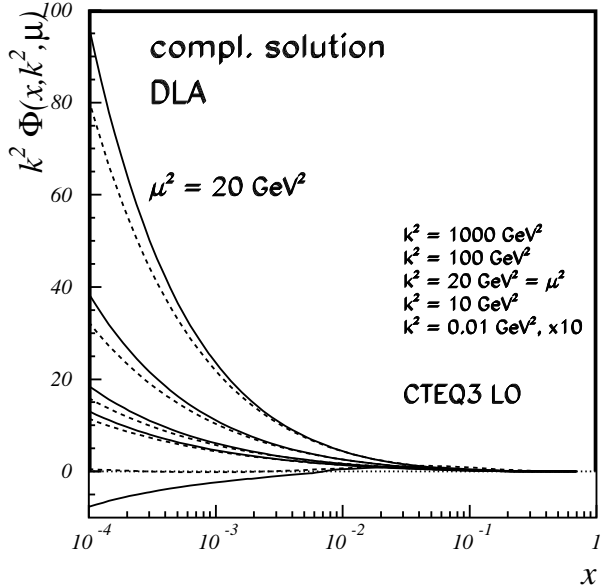

**Fig. 2** The $k_\perp$ dependent gluon distribution as a function of $k^2$ and $x$. Full lines: complete solution eq. (5); dashed lines: solution in DLA. For the input distribution $g(x,\mu)$ result for $k^2 \gtrsim \mu^2$ at $x \lesssim 10^{-3}$ by 10 to 15% while for $k^2 \to 0$ smaller values are obtained. At larger values of $x$ the complete solution approaches the DLA result. For $k^2 \to 0$ $\Phi(x, k^2, \mu)$ vanishes. Since the DLA result is proportional to $J_0(2\sqrt{\overline{\alpha}_s \log(1/x) \log|k^2/\mu^2|})$ for $k^2 \to 0$ a damped oscillatory behaviour is obtained in this approximation. The complete solution, on the other hand, behaves monotonous in the whole kinematical range.

The calculation of $\mathcal{G}(x, k^2, \mu)$ in eq. (5) by numerical Mellin inversion using the numerical solution of eq. (4) is rather time consuming compared to the convolution of the analytical solution (sect. 3) with the input distribution $g(x, \mu)$. We compared both methods and found that the representation given in the previous section leads to a relative error of less than 0.002 using an expansion up to $O(\overline{\alpha}_s^{20})$.

Since shape and size of the complete solution and in DLA are rather similar very precise measurements are required to establish the non–DLA contributions at small $x$.

## 5. Conclusions

We have calculated the $k_\perp$ dependent gluon density numerically in leading order using the BFKL equation. A consistent treatment of observables is possible in the scheme [2, 3]. The Green's function $\mathcal{G}(x, k^2, \mu)$ was found both numerically and by a perturbative analytic expression expanding the complete solution up to $O(\overline{\alpha}_s^{20})$. Both representations agree better than 0.002 after convoluting with the respective input distributions $g(x, \mu)$.

The effect of the non–DLA terms in $\Phi(x, k^2, \mu)$ is of $O(10...15\%)$. To reveal these contributions requires very accurate measurements in the small $x$ range in the case of *every* observable being sensitive to the gluon density.